# Long-range Single Molecule Förster Resonance Energy Transfer Between Alexa Dyes in Zero-Mode Waveguides

Mikhail Baibakov, Satyajit Patra, Jean-Benoît Claude, Jérôme Wenger*

*Aix Marseille Univ, CNRS, Centrale Marseille, Institut Fresnel, 13013 Marseille, France*

* Corresponding author: jerome.wenger@fresnel.fr

**Abstract**

Zero-mode waveguides (ZMW) nanoapertures milled in metal films were proposed to improve the FRET efficiency and enable single molecule FRET detection beyond the 10 nm barrier, overcoming the restrictions of diffraction-limited detection in a homogeneous medium. However, the earlier ZMW demonstrations were limited to the Atto 550 – Atto 647N fluorophore pair, asking the question whether the FRET enhancement observation was an artefact related to this specific set of fluorescent dyes. Here, we use Alexa Fluor 546 and Alexa Fluor 647 to investigate single molecule FRET at large donor-acceptor separations exceeding 10 nm inside ZMWs. These Alexa fluorescent dyes feature a markedly different chemical structure, surface charge and hydrophobicity as compared to their Atto counterparts. Our single molecule data on Alexa 546 – Alexa 647 demonstrate enhanced FRET efficiencies at large separations exceeding 10 nm, extending the spatial range available for FRET and confirming the earlier conclusions. By showing that the FRET enhancement inside a ZMW does not depend on the set of fluorescent dyes, this report is an important step to establish the relevance of ZMWs to extend the sensitivity and detection range of FRET, while preserving its ability to work on regular fluorescent dye pairs.

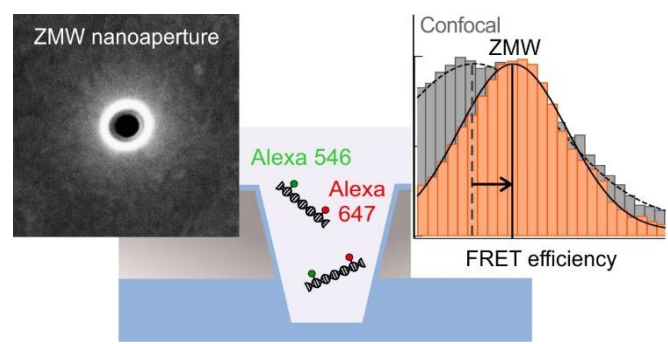

Single molecule Förster resonance energy transfer (smFRET) is a highly sensitive approach to investigate intra and inter-molecular distances on the nanometer scale,[1] revealing dynamic information about biomolecular structures and interactions.[2,3] However, the energy transfer efficiency quickly vanishes when the donor-acceptor separation grows, making smFRET measurements highly challenging at distances above 10 nm.[4,5] Extending smFRET to large biomolecular constructs requires the use of elaborated donor-acceptor constructs, and several strategies have been investigated using lanthanides,[6–8] quantum dots,[9,10] multi-color cascaded systems,[11,12] gold nanoparticles quenchers,[13–15] metal-induced energy transfer,[16,17] or multiple fluorophores.[18–20] Because they do not rely on the organic fluorescent dyes pairs conventionally used in smFRET like Cy3-Cy5 for instance, these advanced approaches further complicate the sample preparation and data analysis. For many applications it would be desirable to extend the smFRET range using regular fluorophore pairs.

Since the works by Purcell and Drexhage,[21,22] it is established that the fluorescence emission decay rate is not determined only by the molecular structure, but also depends on the photonic environment surrounding the molecule. The presence of a mirror (or a more elaborated optical component) can affect the fluorescence decay kinetics and the fluorescence lifetime. In a conceptually similar fashion, the dipole-dipole interaction leading to FRET can also be influenced by the photonic environment in some cases.[23–26] This opens a broad field of research using mirrors,[27–31] microcavities,[23,26,32,33] nanoapertures,[34–40] nanoparticles,[41–48] nanogap antennas,[49–53] or hyperbolic metamaterials.[54,55] Tuning FRET with nanophotonics can potentially overcome the 10 nm barrier in diffraction-limited confocal microscopes while still using conventional fluorophore pairs. However, reaching an enhancement of the FRET efficiency requires a delicate balance between the FRET rate and the other donor radiative and non-radiative processes,[47,48,51,55] while in many cases the FRET efficiency can end up being quenched by the nanophotonic element.[27,33,48,50,51]

We have recently shown that nanoapertures milled in an opaque aluminum film (so called zero-mode waveguides ZMWs[56,57]) can improve the FRET efficiency and enable smFRET detection beyond the 10 nm barrier.[58] ZMWs are promising devices to perform smFRET on large biomolecular constructs with conventional dyes. However, the demonstration was so far limited to Atto 550 – Atto 647N FRET pairs.[58] Both of these fluorescent molecules bear a positive charge after DNA labelling, and have been found to be quite hydrophobic.[59,60] They bear also a higher affinity for glass or metal surfaces, which was observed for Atto 550 and Atto 647N dyes as compared to their cyanine or Alexa Fluor counterparts.[59–62] Although care was taken in our previous work to properly passivate the ZMW surface,[58,62] we cannot fully exclude that the observed enhanced FRET could be related to this specific choice of FRET pair from Atto dyes.

Here, we build on the methodology previously developed for smFRET inside a ZMW,[58] and explore the enhancement of smFRET efficiency for the Alexa Fluor 546 – Alexa Fluor 647 donor-acceptor pair. This FRET pair, although spectrally quite similar to the Atto 550 – Atto 647N pair, has a markedly different chemical structure and behavior (Fig. 1a). Both Alexa 546 and Alexa 647 feature a negative charge once labeled to DNA, while the Atto 550 and 647N dyes have a positive charge. It was observed that these Alexa dyes are more hydrophilic than their Atto counterparts,[59,60] and that the Atto 550 and 647N could induce surface adhesion of the DNA molecules while the Alexa 546 and 647 did not.[62] To assess the relevance of ZMWs for smFRET enhancement, it is thus necessary to quantify their performance for a clearly different set of dyes than the Atto 550 – Atto 647N pair used so far.[35,36,50,58] Our new measurements for Alexa 546 – Alexa 647 smFRET inside aluminum ZMWs demonstrate enhanced smFRET efficiencies at separations exceeding 10 nm, confirming the earlier conclusions drawn with the Atto dyes. The detailed characterization reported here is an important step to establish the relevance and validity of ZMWs to extend the FRET detection range. As additional advantage, all the smFRET measurements in the ZMWs are performed at 100 nM concentration, which is a thousand-fold more concentrated than the conditions typically used for confocal detection. This brings smFRET analysis closer to physiological concentrations.[63,64]

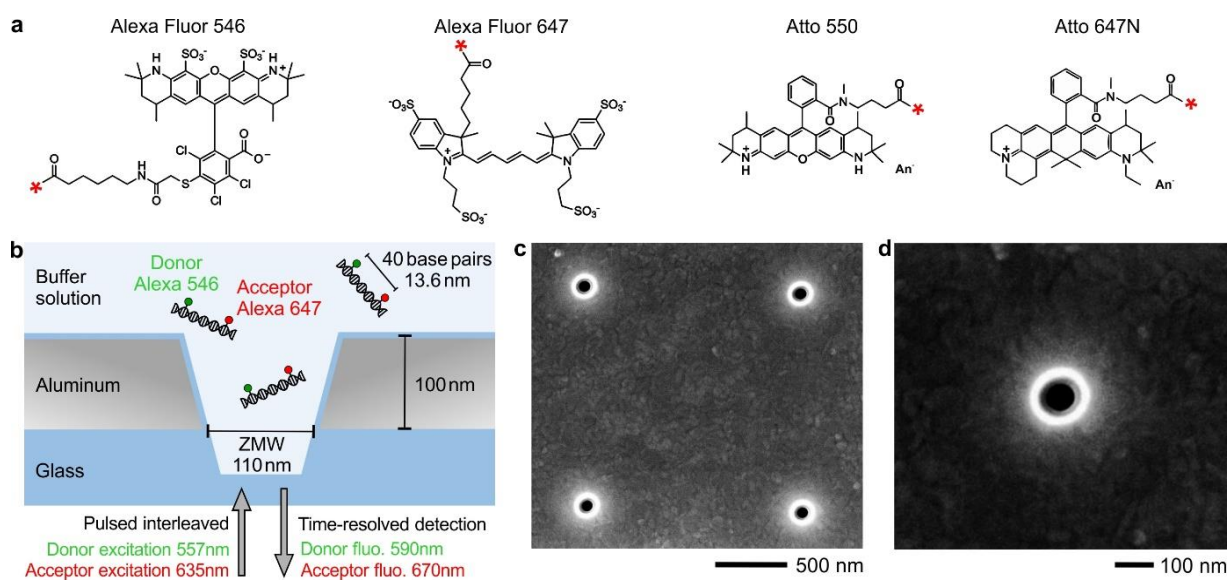

**Figure 1.** (a) Chemical structures of Alexa and Atto fluorescent molecules used as FRET pairs. The red star indicates the DNA labeling site. (b) Experimental scheme of double stranded DNA molecules containing a single Alexa Fluor 546 (donor) and Alexa Fluor 647 (acceptor) FRET pair. The DNA is free to diffuse across the zero-mode waveguide (ZMW) volume where it experiences pulsed interleaved excitation with alternating green and red laser pulses. (c,d) The scanning electron microscope (SEM) images of the pattern with ZMWs and a single ZMW of 110 nm diameter milled in an aluminum film.

**Results and Discussion**

The FRET sample consists of double stranded DNA molecules of 51 base pair length, labeled with a single Alexa 546 donor and a single Alexa 647 acceptor. The donor-acceptor distance is fixed to 30 or 40 base pairs depending on the DNA construct (see Methods section for the detailed DNA sequences and sample preparation). These fluorescent dyes feature a different surface charge and hydrophobicity as compared to Atto 550 and Atto 647N. Alexa 546 and Alexa 647 bear a negative charge after covalent linking to DNA, whereas Atto 550 and 647N have a positive charge (Fig. 1a). A quantitative distinction between the hydrophobicity found for Alexa and Atto dyes can be done by comparing their distribution coefficient logD, with D denoting the ratio of the solute concentration in a nonpolar and a polar solvent. Positive values of logD indicate hydrophobicity (Atto 550 and Atto 647N have logD values of 6.41 and 3.26 respectively), while negative values demonstrate hydrophilicity (Alexa 546 and Alexa 647 have logD of -1.43 and -4.26).[60]

Our experiments monitor the FRET events stemming from individual molecules diffusing across the detection volume (Fig. 1b). To clearly quantify the FRET efficiency and avoid the issues related to incomplete fluorophore labelling, we implement pulsed interleaved excitation (PIE) using two alternating laser excitations to excite the donor and the acceptor dyes in a sequential manner.[65,66] PIE allows to post-select the events corresponding to an active FRET pair where both dyes are fluorescent, and discard all the case where only the donor is present.

The main difference as compared to a conventional diffraction-limited microscope is the use of zero-mode waveguide (ZMW) nanoapertures to confine the light into attoliter volumes.[56,57] The ZMWs used here are milled in a 100 nm thick aluminum film with a diameter of 110 nm (Fig. 1c,d). While the use of Atto 550 – Atto 647N dyes requires the ZMW to be passivated with a silane-modified polyethylene glycol in order to avoid surface adsorption of the DNA molecules,[62] for Alexa 546 – Alexa 647 we find that similar results can be obtained with and without the surface passivation step. This additional advantage of the Alexa FRET pair further simplifies the experiment preparation.

Figure 2 shows typical fluorescence time traces recorded with the confocal setup and with a 110 nm diameter ZMW. In order to ensure that the fluorescence bursts correspond to single molecules passing through the observation volume and to have a negligible probability to observe more than one molecule, we use a low concentration of the DNA sample: 100 pM for confocal and 100 nM for ZMW. Brighter detection events are directly obtained with the ZMW (Fig. 2d-f) as compared to the confocal reference (Fig. 2 a-c), which illustrates one specific advantage of the fluorescence enhancement occurring inside ZMWs as an improvement for the net detected fluorescence brightness.[67] We analyze the total fluorescence time trace using fluorescence correlation spectroscopy (FCS) to compute the temporal autocorrelation and estimate the average number of emitters and their average brightness. Following our previous studies,[58,67] this quantifies the fluorescence brightness enhancement factor for isolated donor and acceptor molecules. For Alexa 546, we find a gain of 11.4 ± 0.9 in a 110 nm ZMW, while for Alexa 647, the enhancement is 15.9 ± 1.2. The fact that a higher enhancement is observed for the red dye is mostly related to the lower quantum yield of the dye (33% for Alexa 647 and 79% for Alexa 546), as low quantum yield emitters lead to the observation of higher enhancement factors.[68]

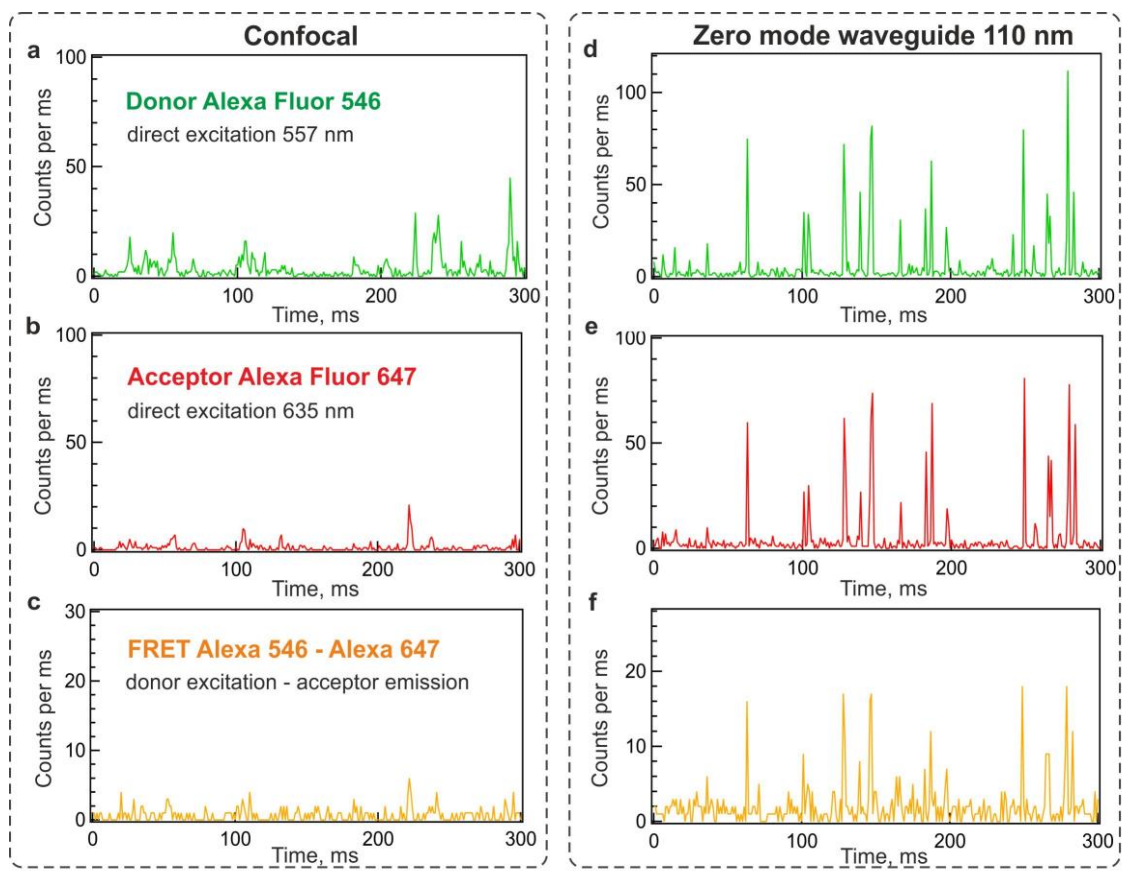

**Figure 2.** Fluorescent time traces of single Alexa546-Alexa647 FRET pairs with 13.6 nm (40 base pairs) separation diffusing in the confocal setup (a-c) and in a 110 nm diameter ZMW (d-f) with 0.5 ms binning time. The traces (a,d) show the donor emission after donor direct excitation at 557 nm, the traces (b,e) show the acceptor emission after acceptor direct excitation at 635 nm, and the traces (c,f) show the FRET emission (acceptor fluorescence) after donor excitation at 557 nm. The fluorescence enhancement in the ZMW directly leads to brighter detection events (d-f) as compared to the confocal reference (a-c) without any post-processing.

Using these fluorescence time traces (the total length is 120 s and accumulates over 2000 detection events), we apply the standardized smFRET analysis protocol detailed in Ref [1] (see Methods for details). After the PIE post-selection, we compute the FRET efficiency $E_{FRET}$ for each burst, taking into account the donor crosstalk, the acceptor direct excitation and the different quantum yields and detection efficiencies between the dyes. The influence of the ZMW is fully taken into account by calibrating the correction parameters for each ZMW independently. The aluminum ZMWs used here are optically weakly resonant components, which do not noticeably modify the fluorescence spectrum of the dyes. Therefore, as we detail in the Methods section, most correction parameters (for crosstalk and direct excitation) are unchanged in the ZMW as compared to the confocal case.

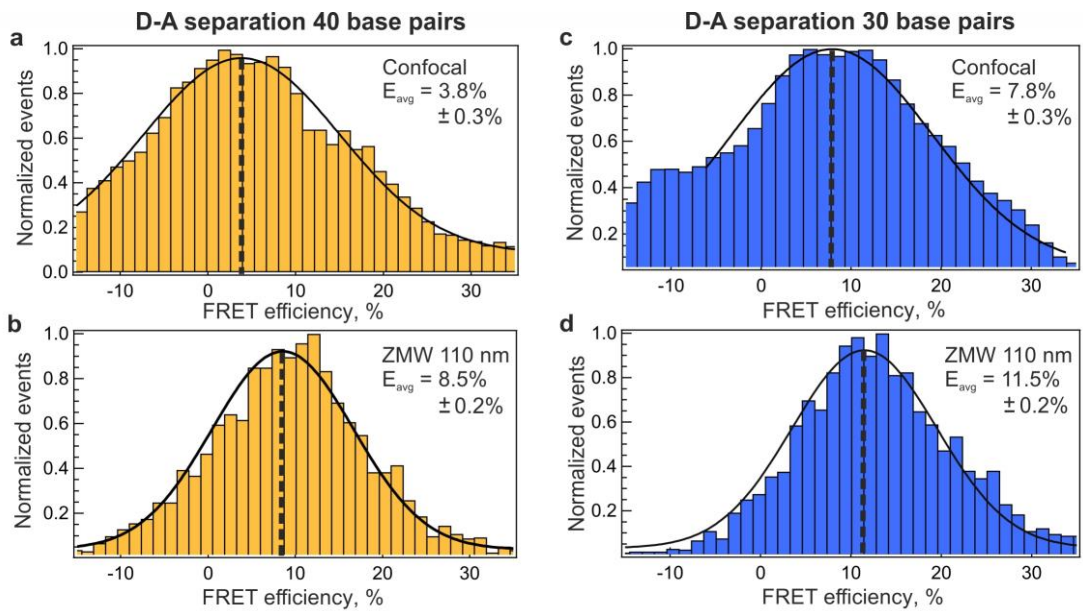

**Figure 3.** Enhancement of the FRET efficiency between Alexa Fluor dyes inside a ZMW. (a,c) smFRET efficiency histograms measured in confocal configuration for Alexa546-Alexa647 FRET pairs with 40 and 30 base pairs respectively (13.6nm and 10.2 nm). (b,d) Same as (a,c) recorded in a ZMW of 110 nm diameter. Black lines are numerical fits with a Gaussian distribution to determine average FRET efficiency (thick dashed line). The average FRET efficiency is indicated for each plot, and the error bar corresponds to one standard deviation of the average value estimate.

Figure 3a,b compares the smFRET efficiency histograms for the Alexa 546 – Alexa 647 construct with 40 base pair separation (corresponding to an average D-A distance of about R = 13.6 nm) for the confocal setup and the ZMW. For this D-A separation, which is about twice the $R_0$ = 7.4 nm Förster radius for this FRET pair, the average FRET efficiency in the confocal case is only 3.8 % ± 0.3 %. To estimate the uncertainty $\sigma_{average}$ on the average FRET efficiency, we apply the classical formula $\sigma_{average} = \sigma/\sqrt{N}$, where $\sigma$ is the standard deviation of the Gaussian distribution fit and $N$ the total number of detected bursts (typically 2000). Using the Förster formula $1/(1+(R/R_0)^6)$ gives a 2.5% estimate for the average FRET efficiency in the confocal case for the 40 base pairs separation. However, this approach is limited by the uncertainties on both R and $R_0$ and the assumption of perfectly random orientation for both dyes, which may not be fully verified by our real sample. Thanks to the optical confinement occurring in the ZMW, the FRET efficiency is improved up to 8.5 % ± 0.2 % inside the 110 nm ZMW. Moreover, the full statistical distributions are clearly different, and smFRET is better detected in the ZMW case thanks to higher average FRET efficiencies and narrower distributions. In the confocal case, the standard deviation of the Gaussian distribution (Fig. 3a,c) is 12%, while it is reduced inside the ZMW to 8% thanks to the higher fluorescence brightness.

We also investigate shorter separations of 30 base pairs (D-A distance ~ 10.2 nm). As the acceptor is brought closer to the donor, the average FRET efficiency is increased to 7.8 % ± 0.3 % in the confocal setup (Fig. 3c) and is further enhanced to 11.5 % ± 0.2 % in the ZMW (Fig. 3d). Computing the gains in the average FRET efficiencies brought by the ZMW, we find a gain of 2.2× (± 0.2) for the 40 bp separation and 1.5x (± 0.1) for the 30 bp case. First, these values demonstrate that the ZMW can indeed improve the net detected FRET efficiency for Alexa FRET pairs, which is especially relevant at large D-A separations exceeding 10 nm where confocal microscopes face their limit of detection. It was observed previously with nanoapertures,[35] nanoantennas,[50] and planar microcavities[26] that the enhancement factors for the FRET rate were higher for the samples corresponding to the larger D-A separations. In other words, the influence of the nanophotonic structure is more pronounced when the D-A separation is larger. We retrieve this feature here. The main reason behind this is that the nanophotonic structure influence on the FRET rate is quite weak as compared to the FRET rate between two closely separated fluorescent dyes in a homogeneous environment. Therefore, one has to go to D-A distances greater than 10 nm so that the ZMW relative influence becomes more prominent.[58] Second, we can compare the results for Alexa and Atto FRET pairs. For Atto 550 – Atto 647N, our previous measurements indicated a gain of 2.9× (± 0.3) for the 40 bp separation and 1.2x (± 0.1) for the 30 bp case,[58] which are quite comparable to the results found here with Alexa dyes despite their different chemical structures. This suggests that the smFRET enhancement inside ZMWs does not depend on the type of fluorescent dyes used.

Independently of the FRET analysis in Fig. 3, the average FRET efficiency can also be assessed from the reduction of the donor lifetime due to the presence of the acceptor, using the formula $E_{FRET} = 1 - \tau_{DA}/\tau_D$, where $\tau_{DA}$ and $\tau_D$ are the fluorescence lifetimes of the donor in presence and absence of acceptor respectively.[2] This approach importantly provides an independent control on the estimated FRET efficiencies, and does not require any separate calibration to account for donor crosstalk, acceptor direct excitation or quantum yield difference. Figure 4a shows the normalized fluorescence decay traces for the Alexa 546 donor in the confocal setup and in a 110 nm diameter ZMW, both with and without an Alexa 647 acceptor. Without any data processing, it is apparent on the fluorescence decays that the presence of the acceptor dye accelerates the donor decay dynamics. This provides a direct evidence for the occurrence of FRET and not of radiative energy transfer mediated by a propagating photon. The donor lifetime change due to the acceptor occurs only in FRET where the dipoles are coupled in the near field *via* evanescent waves. On the contrary, the donor lifetime is unchanged when the dipoles are coupled through radiative transfer, where the energy travels in the form of a propagating photon and can be funneled by the presence of a waveguide.[69–72] The analysis of the traces in Fig. 4a quantifies the $\tau_D$ and $\tau_{DA}$ lifetimes used to compute the average FRET efficiency (Fig. 4b, see details in the Methods section). An excellent agreement is found with the results derived from the smFRET histograms in Fig. 3a,b, with a difference of less than 0.4 % for both the confocal setup and the ZMW. This further confirms the validity of our results.

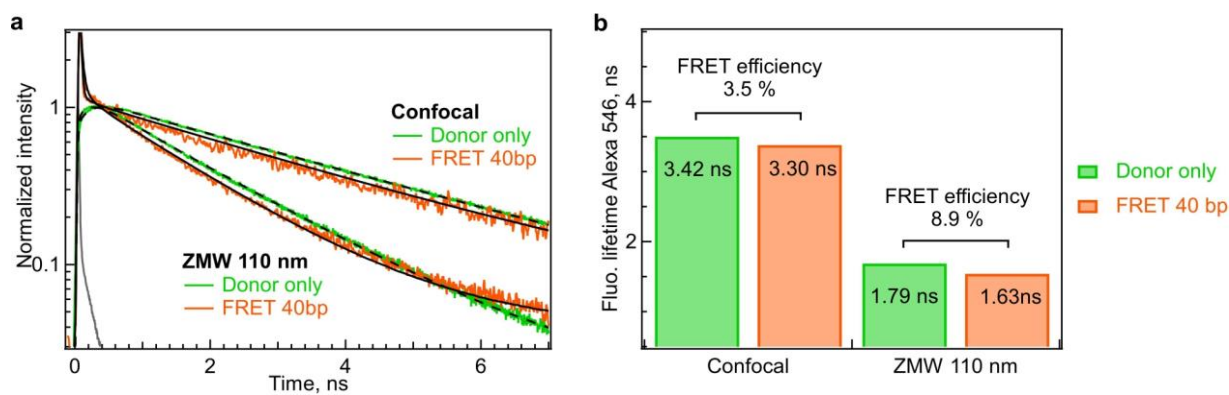

**Figure 4.** (a) Normalized donor fluorescence lifetime decay traces in confocal and in a 110 nm diameter ZMW. Black lines are fits for lifetime traces. For both ZMW and confocal cases, the presence of the acceptor (FRET case) further accelerates the donor decay dynamics, which is a clear signature for FRET. All fit details are summarized in Tab. 1 and in the method section. (b) Intensity-averaged fluorescence lifetimes deduced from the traces in (a), which allow an independent measurement of the average FRET efficiency based on the Alexa 546 donor lifetime reduction.

While the various results demonstrate the FRET enhancement inside a ZMW of 110 nm diameter, we now investigate the influence of the ZMW diameter on the energy transfer between Alexa dyes. The same procedure as for Fig. 3 is applied for ZMW diameters ranging from 80 to 150 nm (Fig. 5a). A gradual shift of the mean FRET efficiency is observed on the distributions, which is summarized as a function of the ZMW diameter in Fig. 5b. Here the 110 nm diameter used in Fig. 3 is near the optimum. Using a large diameter, the fluorescence enhancement decreases and the FRET results tend to retrieve the confocal values. Using a lower diameter, the quenching losses due to direct energy transfer to the metal increase, which compete with FRET to the acceptor dye and reduce the observed FRET efficiency.

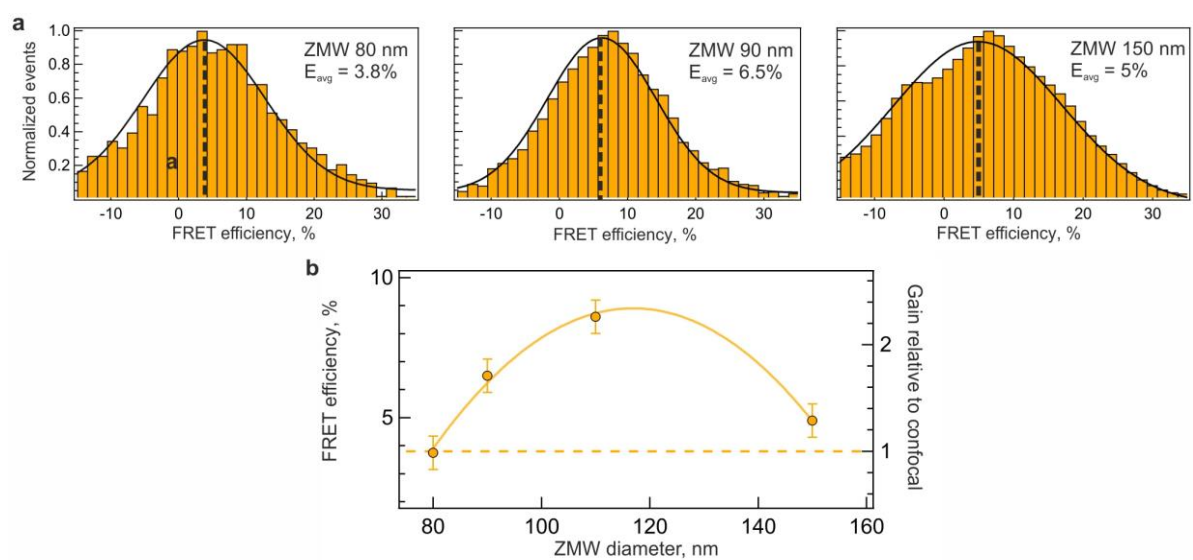

**Figure 5.** (a) smFRET efficiency histograms with ZMWs of different diameters. The DNA sample consists of Alexa546 donor - Alexa647 acceptor with 40 base pair separation similar to Fig. 3b. (b) FRET efficiency (left axis) and FRET efficiency gain (right axis) as a function of the ZMW diameter for Alexa546-Alexa647 with 40 base pair separation. The horizontal dashed line indicates the level found for the confocal reference. The error bar on the graph corresponds to two times the standard deviation on the mean FRET efficiency estimate.

**Conclusions**

Our data demonstrate here that the phenomenon of smFRET enhancement inside a ZMW is quite general and does not depend on the type of fluorescent dyes used. Performing smFRET measurements on Alexa 546 – Alexa 647 pairs, which feature a markedly different chemical structure, surface charge and hydrophobicity as compared to their Atto 550 – Atto 647N counterparts, we retrieve the same conclusions about the quantitative performance of ZMWs to enhance smFRET. Notably, we could achieve over a two-fold enhancement of the net detected FRET efficiency for dyes separated by more than 10 nm. This significantly improves the sensitivity and detection range of smFRET, while preserving the ability to work on conventional fluorophore pairs. The only difference as compared to a classical confocal microscope concerns the replacement of the glass coverslip by a coverslip holding ZMW nanoapertures, which are easy to fabricate using various lithography techniques. The ZMW confine the detection volume to the attoliter range, enabling single molecule FRET detection at a 100 nM concentration. This 1000-fold higher concentration for smFRET than with a diffraction-limited confocal microscope is especially interesting for exploring protein-protein and protein-DNA interactions featuring lower affinities.[63,64] It should be reminded that the Förster formula $1/(1+(R/R_0)^6)$ is only valid for a homogeneous medium and can no longer be applied directly in the vicinity of a nanophotonic structure. Inside a ZMW, both the $R_0$ and the $1/R^6$ decay are affected. While the changes remains minimal for distances below 8 to 10 nm,[35,36] some significant deviations can be found for D-A distances greater than 10 nm, where the FRET enhancement becomes important, as demonstrated in this work. Therefore, preliminary calibrations should be performed in order to enable relevant distance measurements using FRET in ZMWs (and this work contributes to it), but conceptually there is no reason why quantitative distance measurements would not be possible using FRET inside ZMWs.

**Methods**

*Zero-mode waveguide fabrication.* A 100 nm thick layer of aluminum is deposited on a clean glass coverslip by electron-beam evaporation (Bühler Syrus Pro 710).[58] The deposition rate is 10nm/s at a chamber pressure of $5.\,10^{-7}$ mbar.[73] Individual ZMWs are then milled with a gallium-based focused ion beam (FEI dual beam DB 235 Strata) set at 10 pA current and 30 kV voltage. The gallium ion beam has a resolution of about 10 nm. ZMWs are cleaned by UV-ozone during 5 minutes and rinsed with water and ethanol to remove any organic impurities before the measurements.

*Alexa dyes FRET samples.* The FRET sample consists of double stranded DNA with the forward strand being labelled with Alexa Fluor 546 (donor) and its complementary strand with Alexa Fluor 647 (acceptor). The DNA strands are obtained from IBA life solution (Göttingen, Germany) and are HPLC purified. The forward strand sequence of the DNA is 5'-CCT GAG CGT ACT GCA GGA TAG CCT ATC GCG TGT CAT ATG CTG T**T**C AGT GCG-3' where the thymine at position 44 is labelled with Alexa Fluor 546. The complementary reverse strand is 5'-CGC ACT GAA CAG CAT ATG ACA CGC GAT AGG CTA TCC TGC AGT ACG C**T**C AGG-3' where the T base at position 47 is labelled with Alexa Fluor 647. In this configuration, we get a 40 base pairs separation between donor and acceptor dyes corresponding to approximately 13.6 nm. For the sample with 10.2 D-A separation (30 base pairs), the T base at position 37 is instead labelled with Alexa Fluor 647. The forward and reverse strands are hybridized in a buffer containing 5 mM Tris, 20 mM MgCl₂, 5 mM NaCl at pH 7.5. First, the mixture is heated at at 90°C for 5 minutes. Then the mixture is cooled down to room temperature for 3 hours. The concentration of 100 pM and 100 nM is used for the smFRET measurements in confocal and in ZMWs respectively in a buffer containing 20 mM Hepes, 10 mM NaCl, 0.1% Tween 20 at pH 7.5.

*Experimental setup.* The confocal microscope set up has been detailed in Ref. [58]. Briefly, the Alexa 546 donor is excited at 557 nm by a iChrome-TVIS laser (Toptica), and the Alexa 647 acceptor is excited at 635 nm by a LDH laser diode (PicoQuant). Green and red pulses are alternating at 40MHz repetition rate in a PIE configuration,[66,74] with 20 µW average power low enough to avoid fluorescence saturation or photobleaching on the diffusing dyes. Both lasers have linear polarizations which are set parallel to each other. No polarization selection is performed on the fluorescence detection. The microscope objective is a Zeiss C-Apochromat 63x, 1.2 NA water immersion objective used in epifluorescence configuration. For detecting donor and acceptor fluorescence, two MPD-5CTC avalanche photodiodes (Picoquant) are employed together with 50 µm confocal pinholes and spectral filters (donor fluorescence collection from 570 to 620 nm, acceptor fluorescence collection from 655 to 750 nm). The photodiode signals are connected to a HydraHarp400 single photon counting module (Picoquant) in a time-tagged time-resolved (TTTR) mode. The overall system timing resolution is 38 ps (full width at half maximum).

*FRET efficiency measurements.* The procedure to compute the FRET efficiency histograms follows the standard approach in smFRET.[1,74] First, we select the single molecule detection events and separate them from the background noise, applying a threshold criterion so that the sum of the signals in donor and acceptor channels exceeds 25 counts per ms for the ZMW (12 counts per ms for the confocal case). A second threshold is used to check the presence of the red dye upon the excitation by the red laser. We choose the value at 12 counts per ms in acceptor channel with red excitation (3 counts per ms for the confocal case due to the lower fluorescence brightness). We ensure that these levels have a negligible influence on the measured average FRET efficiencies.

The FRET efficiency is then calculated as [1,58]

$$E_{FRET} = \frac{n_A^{green} - \alpha n_D^{green} - \delta n_A^{red}}{\left(n_A^{green} - \alpha n_D^{green} - \delta n_A^{red}\right) + \gamma n_D^{green}} \quad (1)$$

where $n_D^{green}$ and $n_A^{green}$ the number of photons per each burst for donor and acceptor channel upon excitation by a green laser, $n_A^{red}$ number of photons in red channel with the excitation by the red laser.

The numbers of photons are corrected for the background contribution in each channel. The background counts are measured by performing a separate experiment using the buffer solution only for the ZMW or the reference glass coverslip. The equation above contains correction factors for (1) the crosstalk α fraction of the donor emission collected in the acceptor detection channel, (2) the direct excitation δ of the acceptor by the green laser and (3) the correction parameter γ which accounts for the difference of the quantum yields of the dyes ($\phi_d$, $\phi_a$) and their detection efficiencies between channels ($\kappa_d$, $\kappa_a$).

The crosstalk α is the ratio between the donor emission leaking into the acceptor channel as compared to the donor emission in the donor channel. The crosstalk is determined for DNA sample containing only the donor fluorophore: $\alpha = \frac{n_A^{green}}{n_D^{green}}$. For all ZMWs diameters, the crosstalk remains nearly constant α=0.05 with slight variation for 100 and 150 nm at α=0.04. We found a similar value for the confocal setup α=0.05.

The direct excitation δ corresponds to the fraction of the acceptor fluorescence due to direct excitation by the green laser as compared to acceptor emission signal upon the red laser. This parameter is measured when a DNA sample containing only the acceptor dye is excited: $\delta = \frac{n_A^{green}}{n_A^{red}}$. For the confocal reference, we find δ=0.11, which also does not change for ZMWs diameters except for the 80 nm ZMW where we have δ=0.08.

The γ correction factor takes into account the differences in the fluorescence quantum yields ($\phi_d$, $\phi_a$) and the detection efficiencies ($\kappa_d$, $\kappa_a$): $\gamma = \frac{\kappa_A \phi_A}{\kappa_D \phi_D}$. For the Alexa 546-Alexa647 FRET pair in our confocal setup, we compute $\gamma_{conf}$=0.43 ± 0.02 from the knowledge of the fluorescence spectrum and quantum yield of each dye. The photodiode response is also accounted for in the calculation. Alternatively, the γ correction factor can also be estimated from the measured stoichiometry S,[1] and the average fluorescence brightness per molecule $CRM_A^{red}$ and $CRM_{DO}^{green}$ measured by FCS following the approach used in Ref. [58] :

$$\gamma_{conf} = \frac{S}{1-S} \frac{CRM_A^{red}}{CRM_{DO}^{green}} \qquad (2)$$

where $CRM_A^{red}$ is for red excitation of the acceptor dye and $CRM_{DO}^{green}$ is for green excitation of the sample containing only the donor dye. Using the measured values of $CRM_A^{red} = 5000$ counts/s for Alexa 647 DNA, $CRM_{DO}^{green} = 11000$ counts/s for Alexa 546 DNA and $S = 0.46$, we find $\gamma_{conf}$=0.39 ± 0.05, in good agreement with the 0.43 ± 0.02 calculated value.

For the ZMW, γ is modified due to the different enhancement factors of the donor and acceptor and is found as: [35,50,58]

$$\gamma_{ZMW} = \gamma_{conf} \times \frac{EnhCRM_{AO}^{green}}{EnhCRM_{DO}^{green}} = \gamma_{conf} \times \frac{\delta_{ZMW}}{\delta_{conf}} \times \frac{EnhCRM_{AO}^{red}}{EnhCRM_{AO}^{green}} \quad (3)$$

where $EnhCRM_{AO}^{green}$, $EnhCRM_{DO}^{green}$ are the fluorescence enhancement factors of the fluorescence count rate per molecule (CRM) for acceptor-only and donor-only samples upon a green excitation, and $EnhCRM_{AO}^{red}$ is for a red excitation. All the enhancement factors are assessed by fluorescence correlation spectroscopy for each ZMW diameter.[67] We find $\gamma_{ZMW}$ for 80 nm, 90nm, 110nm, 150nm as 0.45, 0.65, 0.6, 0.65 respectively. Except for the 80 nm ZMW (for which a significant fluorescence quenching is found), the γ correction factor does not vary much for ZMW diameters from 90 to 150 nm. Contrarily to the case of Atto550-Atto647N FRET pair,[58] for Alexa546-Alexa647 we find an increase of the γ correction factor in the ZMW as compared to the confocal reference ($\gamma_{ZMW}$~0.65 while $\gamma_{conf}$=0.43). According to Eq. (1), a higher γ value will lead to a decrease of the FRET efficiency (by increasing the denominator in the fraction). The net FRET efficiency enhancement observed inside the ZMW (Fig. 3 and 5) shows that the gain in the acceptor emission $n_A^{green}$ is enough to compensate for the increased γ factor.

*Fluorescence lifetime analysis.* In addition to the fluorescence burst analysis Eq. (1), the average FRET efficiency can be independently determined from the donor fluorescence lifetime data. We use the equation $E_{FRET} = 1 - \tau_{DA}/\tau_D$, where $\tau_{DA}$ and $\tau_D$ are the fluorescence lifetimes of the donor in presence and absence of acceptor respectively. To determine $\tau_{DA}$ and $\tau_D$, we fit the time correlated single photon counting (TCSPC) histograms (Fig. 4a) with a reconvolution taking into account the instrument response function (IRF), whose full width at half maximum was measured to be 38 ps. All the lifetime analysis is performed using the Symphotime 64 software (PicoQuant). For the FRET data in Fig. 4a, we use the same traces as for the intensity burst analysis in Fig. 2,3 leading to an average number of molecules in the detection volume around 0.1 (concentrations of 100 pM and 100 nM for the confocal and ZMW cases). For the donor only TCSPC data, we use 10 times higher concentrations (average number of detected molecules about 1) to achieve a better signal to noise ratio. As a consequence of the higher background contribution in the FRET cases, a peak at t = 0 is seen for the FRET TCSPC data. This peak corresponds to a sum of laser light scattering, metal photoluminescence and Raman scattering. This contribution is interpolated with a fixed 20 ps component (shorter than the 38 ps IRF resolution) to achieve a complete fitting of the TCSPC decay, but this contribution is then discarded for the lifetime analysis as it only corresponds to noise. The background and scattering contributions are recorded by performing a separate experiment in the same conditions using only the buffer in the absence of DNA sample. For the TCSPC analysis, we ensure that more than 92% of the total detected photons are considered for the fits. The data for the confocal reference (FRET and donor only) are fitted with a single exponential decay (excluding the fixed 20 ps contribution from laser scattering). For the ZMWs, a biexponential function with fast and slow components provides a better fit with flat residuals. As observed previously for Atto dyes,[58] the fast component converges towards values around 400 ps for both the FRET and donor-only samples inside a 110 nm ZMW. All the fit parameters are summarized in Table 1. The tail seen for long delay times higher than 5 ns is only due to the background level, there is no supplementary long lifetime. In the ZMW case, the intensity-averaged lifetimes are used to compute the average FRET efficiency. We find empirically that these values provide a better match with the separate burst intensity analysis (Fig. 3) than the amplitude-averaged data. However, our claim of enhanced FRET efficiency in the ZMW is maintained for all the approaches (intensity-averaged, amplitude-averaged or direct comparison between long lifetime components).

**Table 1.** Results obtained from the numerical fit to the TSCPC histograms shown in Fig. 4a. In the case of a biexponential fit, $\tau_1$ and $\tau_2$ are the individual lifetimes of each component and $\alpha_1$ and $\alpha_2$ are their respective normalized amplitudes. $\tau_{amp} = (\alpha_1 \tau_1 + \alpha_2 \tau_2)/(\alpha_1 + \alpha_2)$ is the amplitude-averaged lifetime, while $\tau_{int} = (\alpha_1 \tau_1^2 + \alpha_2 \tau_2^2)/(\alpha_1 \tau_1 + \alpha_2 \tau_2)$ denotes the intensity-averaged lifetime. The 20 ps scattering peak at t = 0 is not shown here.

| Condition | Sample | $\tau_1$/ns | $\tau_2$/ns | $\alpha_1$ | $\alpha_2$ | $\tau_{int}$/ns | $\tau_{amp}$/ns |
|---|---|---|---|---|---|---|---|
| **Confocal** | D only | 3.42 | - | 1 | - | **3.42** | 3.42 |
|  | D-A 40bp | 3.30 | - | 1 | - | **3.30** | 3.30 |
| **ZMW 110nm** | D only | 1.89 | 0.40 | 0.76 | 0.24 | **1.79** | 1.54 |
|  | D-A 40bp | 1.77 | 0.40 | 0.67 | 0.33 | **1.63** | 1.32 |

**Notes** The authors declare no competing financial interest.

**Acknowledgments**

The authors thank Antonin Moreau and Julien Lumeau for help with the aluminum layer preparation. This project has received funding from the Agence Nationale de la Recherche (ANR) under grant agreement ANR-17-CE09-0026-01 and from the European Research Council (ERC) under the European Union's Horizon 2020 research and innovation programme (grant agreement No 723241).